\documentclass[sigconf]{acmart}

\AtBeginDocument{%
	\providecommand\BibTeX{{%
			\normalfont B\kern-0.5em{\scshape i\kern-0.25em b}\kern-0.8em\TeX}}}

\newcommand{\figref}[1]{Fig.~\ref{#1}}
\newcommand{\secref}[1]{Sec.~\ref{#1}}
\copyrightyear{2019} 
\acmYear{2019} 
\setcopyright{rightsretained} 
\acmConference[DEBS '19]{DEBS '19: The 13th ACM International Conference on Distributed and Event-based Systems}{June 24--28, 2019}{Darmstadt, Germany}
\acmBooktitle{DEBS '19: The 13th ACM International Conference on Distributed and Event-based Systems (DEBS '19), June 24--28, 2019, Darmstadt, Germany}\acmDOI{10.1145/3328905.3332506}
\acmISBN{978-1-4503-6794-3/19/06}

\begin{document}

%
\title{Poster: Benchmarking Financial Data Feed Systems}
\titlenote{This is the author's version of the accepted work posted here for your personal use. Not for redistribution. Definitive Version of Record published in {DEBS '19} (see below).}

\author{Manuel Coenen}
\author{Christoph Wagner}
\authornote{Both authors contributed in equal terms to this research.}
\affiliation{%
	\institution{vwd: Vereinigte Wirtschaftsdienste GmbH}
	\city{Kaiserslautern}
	\country{Germany}
}
\email{wrench@vwd.com}

\author{Alexander Echler}
\author{Sebastian Frischbier}
\affiliation{%
	\institution{vwd: Vereinigte Wirtschaftsdienste GmbH}
	\city{Frankfurt a.M.}
	\country{Germany}
}
\email{wrench@vwd.com}

\renewcommand{\shortauthors}{Coenen and Wagner et al.}

%
\begin{abstract}
Data-driven solutions for the investment industry require event-based backend systems to process high-volume financial data feeds with low latency, high throughput, and guaranteed delivery modes.

At vwd we process an average of 18 billion incoming event notifications from 500+ data sources for 30 million symbols per day and peak rates of 1+ million notifications per second using custom-built platforms that keep audit logs of every event.

We currently assess modern open source event-processing platforms such as Kafka, NATS, Redis, Flink or Storm for the use in our ticker plant to reduce the maintenance effort for cross-cutting concerns and leverage hybrid deployment models. For comparability and repeatability we benchmark candidates with a standardized workload we derived from our real data feeds.

We have enhanced an existing light-weight open source benchmarking tool in its processing, logging, and reporting capabilities to cope with our workloads. The resulting tool \emph{wrench} can simulate workloads or replay snapshots in volume and dynamics like those we process in our ticker plant. We provide the tool as open source.

As part of ongoing work we contribute details on (a) our workload and requirements for benchmarking candidate platforms for financial feed processing; (b) the current state of the tool \emph{wrench}.

\end{abstract}



%

%
\maketitle

%
%

\section{Introduction}

Event-processing platforms are the backbone of any data-driven solution favoured by the investment industry. Participants of today’s financial markets require accurate information about meaningful events delivered to them in a timely manner based on their individual quality of information (QoI) and quality of service (QoS) requirements. Thus, event-processing platforms for financial data feeds have to simultaneously excel in multiple functional and non-functional aspects such as throughput, latency, order, completeness, and availability.  Furthermore, they have to keep audit-safe logs of every processed notification for regulatory and compliance reasons.

As one of Europe’s leading providers of financial data and regulatory solutions, vwd operates extensive custom-built event-driven platforms on dedicated infrastructure to process and enrich high-volume streams of financial data at low latency. Our production systems process an average of 18 billion incoming event notifications for 30 million symbols per day. The data received from our 500+ data sources is purged, enriched, and normalised within our ticker plant and delivered to our internal and external subscribers in a proprietary condensed binary format.  While our closed source publish/subscribe and complex event-processing (CEP) systems are tailored to the characteristics of financial data streams and specific use cases, their tight vertical integration hampers the use in hybrid environments as these rely on horizontal scalability and isolation. 

Modern open source event-processing systems like Kafka, NATS, Redis, Flink or Storm in turn are designed to run on highly scalable platforms like Kubernetes. They have emerged from various communities to cover a wide spectrum of use cases and data formats, ranging from log processing in social media to sensor data management in the Internet of Things (IoT). 
However, as multi-purpose platforms they can deliberately balance versatility with degradations in throughput, availability, completeness or latency.

As part of ongoing work we assess such open systems for the use as \emph{event bus platforms} in our ticker plant. For this we want to apply a standardized benchmark to transparently gauge and compare candidates in a repeatable manner. In this contribution we describe the key requirements of our industry for an event bus (\secref{sec:req}), the workload we use for our tests (\secref{sec:workl}), and the current state of the tool (\emph{wrench}) we provide as open source (\secref{sec:tool}).

\section{Related Work}

Designing and implementing representative benchmarks for event-processing platforms remains an active area of research. Most work focuses on identifying performance metrics of CEP systems while few tools actually implement these metrics~\cite{7592714}; of these, micro-benchmarks or domain-specific and scenario-based benchmarks prevail that aim at fine-tuning parameters of a specific system~\cite{shukla2016benchmarking}. 
The Pairs benchmark~\cite{Mendes:2013:TSE:2479871.2479913} is the most promising candidate as it uses an investment strategy scenario. 
However, it focuses on the correctness of the CEP engine's reasoning and uses an oversimplified workload;
its implementation is not fully available and those components that are have been discontinued since 2013.

\section{Investment Industry Requirements}
\label{sec:req}

Event bus solutions to serve in our ticker plant have to satisfy the following mandatory (M) or optional (O) requirements regarding QoS and QoI, workload processing, and architecture: 
 
\emph{(1)~QoS and QoI:} guaranteed end-to-end processing latency of less than 20~ms with exactly-once and order-preserving delivery mode (M); prioritisation (O), i.e., delayed delivery of low-prio notifications allowed to guarantee timely processing of high-prio ones.
\emph{(2)~Workload processing:} ability to deal with peak rates between 700,000 to 1.1 million notifications/sec while the notification size varies between 80 bytes (tick) and 31kb (news) (M); provide data compression for efficiency (O).
\emph{(3)~Architecture}: horizontally scaling distributed system with fail-over and auto-recovery capabilities; logging of all processed notifications with timestamps (send/received) and provenance without performance degradations (M).

Suitable benchmarking tools to assess candidates accordingly must meet at least the following requirements:

 \emph{(A) Processing:} scalability of publishers/subscribers to generate and process described workloads or replay real data snapshots; support custom binary and textual formats; generate/replay workloads varying over time using defined rate distribution patterns.
 \emph{(B) Reporting and analytics:} complete log of latencies for reporting and export to open formats, e.g., comma separated value (CSV).

\section{A Representative Workload}
\label{sec:workl}
 
\begin{figure}[ht]
	\includegraphics[width=5.5cm]{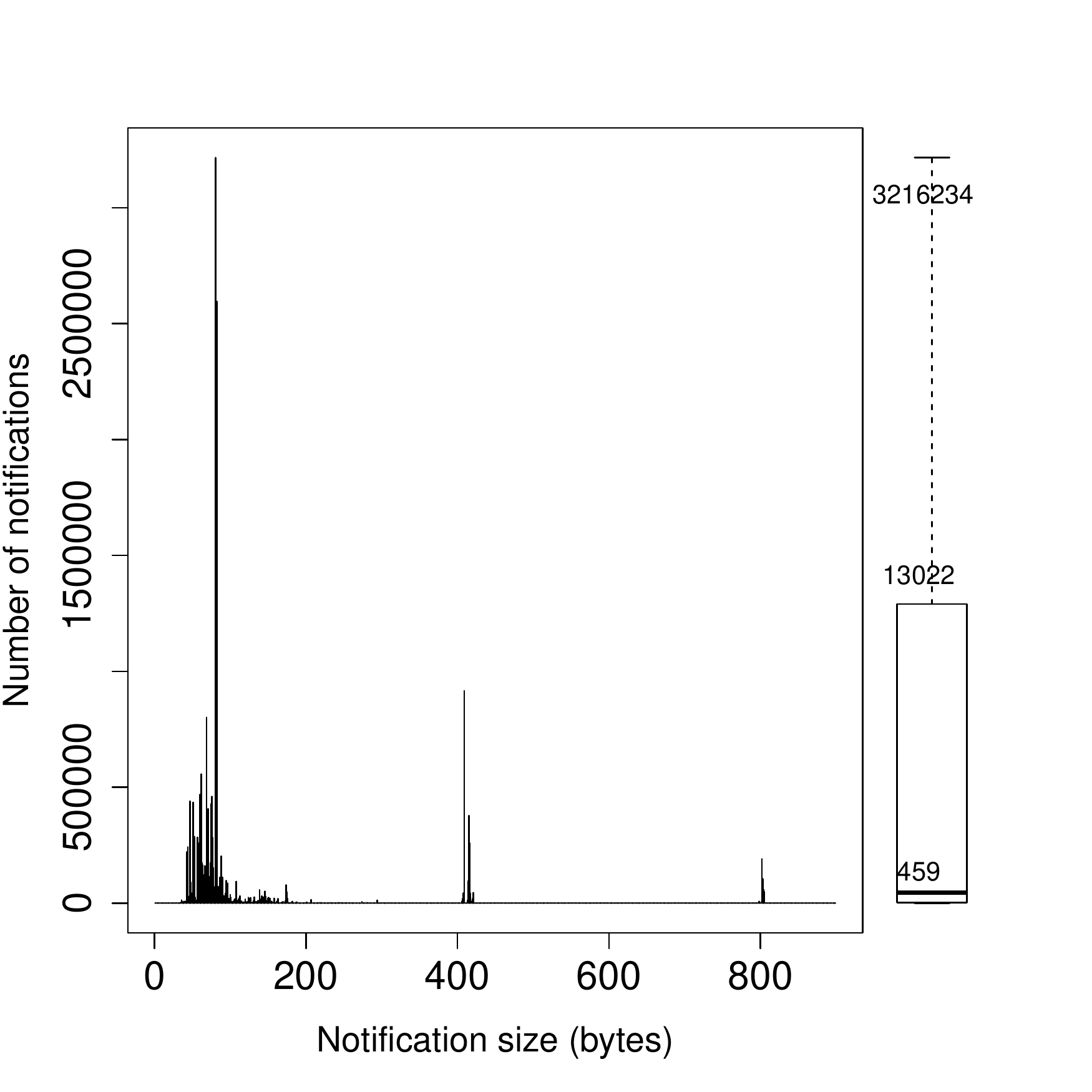}		
	\caption{Workload: notification size distribution}
	\label{fig:dfeed-size-distribution}
\end{figure}

\begin{figure}[ht]
	\includegraphics[width=5.5cm]{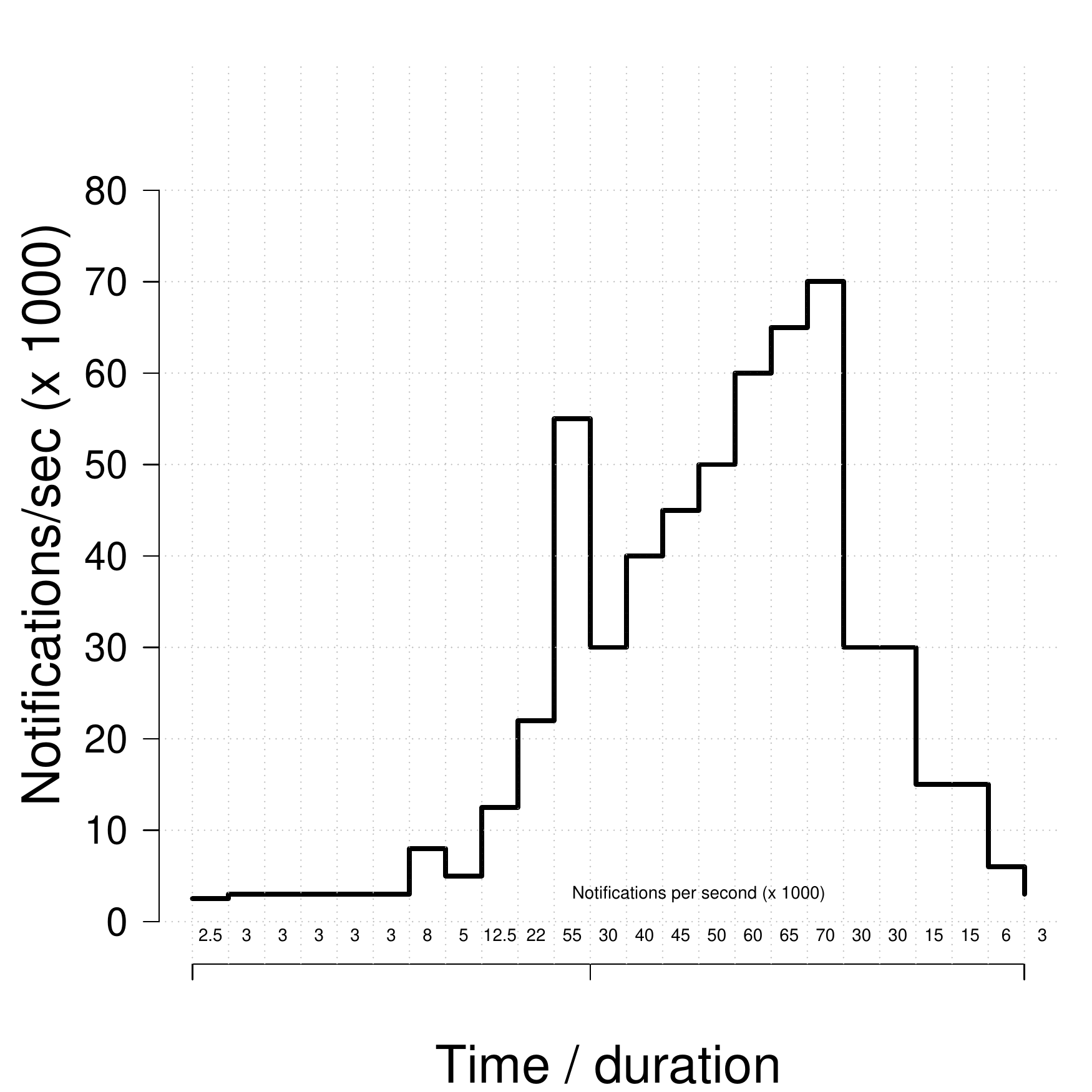}		
	\caption{Workload: scalable input rate profile}
	\label{fig:dfeed-distribution}
\end{figure}

We define a baseline workload using a generalized input rate profile and data of a 60 seconds snapshot we took from one of our normalized internal feeds during a representative trading day.  
The snapshot contains 18,023,662 notifications of 8 event types (no news) with 6 to 129 attributes each (median: 16) about 2.89 million symbols totalling 2.3 GB. \figref{fig:dfeed-size-distribution} shows the size of notifications, omitting < 50 notifications larger than 900 bytes. \figref{fig:dfeed-distribution} shows the input rate profile generalized from a typical 24h input rate distribution. 
For benchmarking we scale rates and volume with varying parameters to simulate different scenarios, e.g., Brexit.

\section{Tool Support: Enhancing bench}
\label{sec:tool}

After reviewing available tooling we decided to enhance the existing light-weight open source benchmarking tool \emph{bench}~\cite{ttbench}.  
Built in Go with native support for NATS (streaming), Kafka, Cassandra, and Redis to tune their setup, its unmodified version is too limited regarding processing and reporting for our purpose. We provide the enhanced tool \emph{wrench} on GitHub~\cite{fdbench} (work in progress).

\emph{Processing capabilities:} so far we split publishers and consumers into separate implementations for increased horizontal scalability and improved bench's throttling algorithm so that our high-volume workloads can be simulated. We improved the expressiveness of workload definitions so that more complex rate distributions can be modelled and simulated with random values or by replaying existing snapshots from custom data formats: input rate profiles for publishers can be provided in CSV format by defining durations in seconds with a corresponding load (e.g.,~\figref{fig:dfeed-distribution}); the dimensions can be scaled by a parameter. More complex subscription schemes and simulating churn are features currently being worked on.

\emph{Analytical capabilities:} raw latency information is now stored in a binary format, allowing wrench to capture information about larger datasets without  performance degradations. A converter tool to a CSV format for post-processing is provided accordingly.

\bibliographystyle{ACM-Reference-Format}
\bibliography{ManuelCoenen}


\begin{thebibliography}{5}


\ifx \showCODEN    \undefined \def \showCODEN     #1{\unskip}     \fi
\ifx \showDOI      \undefined \def \showDOI       #1{#1}\fi
\ifx \showISBNx    \undefined \def \showISBNx     #1{\unskip}     \fi
\ifx \showISBNxiii \undefined \def \showISBNxiii  #1{\unskip}     \fi
\ifx \showISSN     \undefined \def \showISSN      #1{\unskip}     \fi
\ifx \showLCCN     \undefined \def \showLCCN      #1{\unskip}     \fi
\ifx \shownote     \undefined \def \shownote      #1{#1}          \fi
\ifx \showarticletitle \undefined \def \showarticletitle #1{#1}   \fi
\ifx \showURL      \undefined \def \showURL       {\relax}        \fi
\providecommand\bibfield[2]{#2}
\providecommand\bibinfo[2]{#2}
\providecommand\natexlab[1]{#1}
\providecommand\showeprint[2][]{arXiv:#2}

\bibitem[\protect\citeauthoryear{Gradvohl}{Gradvohl}{2016}]%
        {7592714}
\bibfield{author}{\bibinfo{person}{Andre~L.S. Gradvohl}.}
  \bibinfo{year}{2016}\natexlab{}.
\newblock \showarticletitle{Investigating Metrics to Build a Benchmark Tool for
  Complex Event Processing Systems}. In \bibinfo{booktitle}{\emph{IEEE 4th
  International Conference on Future Internet of Things and Cloud Workshops}}
  \emph{(\bibinfo{series}{FiCloudW '16})}. \bibinfo{publisher}{IEEE},
  \bibinfo{pages}{143--147}.
\newblock


\bibitem[\protect\citeauthoryear{Mendes, Bizarro, and Marques}{Mendes
  et~al\mbox{.}}{2013}]%
        {Mendes:2013:TSE:2479871.2479913}
\bibfield{author}{\bibinfo{person}{Marcelo~R.N. Mendes}, \bibinfo{person}{Pedro
  Bizarro}, {and} \bibinfo{person}{Paulo Marques}.}
  \bibinfo{year}{2013}\natexlab{}.
\newblock \showarticletitle{Towards a Standard Event Processing Benchmark}. In
  \bibinfo{booktitle}{\emph{Proceedings of the 4th ACM/SPEC International
  Conference on Performance Engineering}} \emph{(\bibinfo{series}{ICPE '13})}.
  \bibinfo{publisher}{ACM}, \bibinfo{pages}{307--310}.
\newblock


\bibitem[\protect\citeauthoryear{Shukla and Simmhan}{Shukla and
  Simmhan}{2017}]%
        {shukla2016benchmarking}
\bibfield{author}{\bibinfo{person}{Anshu Shukla} {and} \bibinfo{person}{Yogesh
  Simmhan}.} \bibinfo{year}{2017}\natexlab{}.
\newblock \showarticletitle{Benchmarking Distributed Stream Processing
  Platforms for IoT Applications}. In \bibinfo{booktitle}{\emph{Performance
  Evaluation and Benchmarking. Traditional - Big Data - Internet of Things}}
  \emph{(\bibinfo{series}{TPCTC '16})}. \bibinfo{publisher}{Springer},
  \bibinfo{pages}{90--106}.
\newblock
\showISBNx{978-3-319-54334-5}


\bibitem[\protect\citeauthoryear{Treat}{Treat}{2017}]%
        {ttbench}
\bibfield{author}{\bibinfo{person}{Tyler Treat}.}
  \bibinfo{year}{2017}\natexlab{}.
\newblock \bibinfo{title}{bench - A generic latency benchmarking library.}
\newblock \bibinfo{howpublished}{\url{https://github.com/tylertreat/bench}}.
\newblock
\newblock
\shownote{[Online; accessed 2019-04-17].}


\bibitem[\protect\citeauthoryear{vwdsrc}{vwdsrc}{2019}]%
        {fdbench}
\bibfield{author}{\bibinfo{person}{vwdsrc}.} \bibinfo{year}{2019}\natexlab{}.
\newblock \bibinfo{title}{wrench - Workload-optimized \& Reengineered bench}.
\newblock \bibinfo{howpublished}{\url{https://github.com/vwdsrc/wrench}}.
\newblock
\newblock
\shownote{[Online; accessed 2019-04-17].}


\end{thebibliography}

\end{document}